\documentclass[prb,showpacs,amsmath,amssymb]{revtex4}
\usepackage{graphicx}% Include figure files
\usepackage{bm}% bold math

\def\br{ \bm{r} }
\def\bk{ \bm{k} }

\def\Tr{ \,\mathrm{Tr}\,}
\def\bq{ \bm{q} }
\def\sgn{ \,\mathrm{sgn}\, }

\begin{document}

\title{Phase solitons and subgap excitations in two-band superconductors}

\author{K. V. Samokhin}

\affiliation{Department of Physics, Brock University,
St.Catharines, Ontario, Canada L2S 3A1}
\date{\today}

\begin{abstract}

A phase soliton is a topological defect peculiar to two-band superconductors, which is associated with a $2\pi$ winding of the relative phase of the two superconducting condensates. 
We study the quasiparticle spectrum in the presence of a single planar phase soliton. We show that the order parameter phase variation in each of the bands leads to the existence of subgap states bound to the soliton. 
Calculation of the soliton energy valid at all temperatures is presented, with exact analytical results obtained for a simple soliton model. 

\end{abstract}


\maketitle

\section{Introduction}
\label{sec: Intro}

The recent resurgence of interest in the properties of multiband, in particular two-band, superconductors has been largely stimulated by the discovery of superconductivity in MgB$_2$ (Refs. \onlinecite{MgB2} and \onlinecite{MgB2-review}). 
Other candidates for multiband superconductivity include nickel borocarbides (Ref. \onlinecite{CGB98}), NbSe$_2$ (Ref. \onlinecite{NbSe2}), the heavy-fermion compounds CeCoIn$_5$ (Ref. \onlinecite{CeCoIn5}) and CePt$_3$Si (Ref. \onlinecite{CePtSi}), and also 
the whole family of iron-based high-temperature superconductors (Ref. \onlinecite{Fe-based}). These discoveries have shown that multiband superconductivity, which is characterized by 
a significant difference in the order parameter magnitudes and/or phases in different bands, is a much more common phenomenon than was previously thought. 

Theoretically, a two-band generalization of the Bardeen-Cooper-Schrieffer (BCS) theory was introduced in Ref. \onlinecite{two-band-BCS}. Subsequent work has shown that many properties 
of multiband superconductors differ qualitatively from the single-band case and that the most spectacular features are associated with the presence of
additional degrees of freedom -- the relative phases of the pair condensates in different bands. For example, in a charged two-band superconductor, the collective mode corresponding to small oscillations of the relative phase, 
called the Leggett mode,\cite{Leggett-mode} is not accompanied by the charge density modulation and, therefore, is not pushed up into the plasma frequency region. 
If the two condensate phases have different windings around the core of a vortex, then the vortex will carry a fractional magnetic flux.\cite{Bab02} 

In addition to exotic vortices, there is another type of topological defects specific to multiband superconductivity, namely phase solitons.\cite{Tanaka02} 
A phase soliton is a topologically stable texture of the superconducting order parameter, in which the relative phase exhibits a kink-like variation by $2\pi$ between its asymptotic mean-field values. 
The phase solitons can be dynamically generated in nonequilibrium current-carrying states,\cite{GV03} or even in static situations by the
proximity effect with a conventional $s$-wave superconductor.\cite{Vakar12} Phase solitons of a different kind, connecting degenerate time-reversal symmetry breaking states, may exist in superconductors with three or more bands.\cite{LH11} 
Stable nontrivial phase textures similar to the phase solitons can also exist in single-band superconductors with unconventional multi-component order parameters. For instance, a chiral $p$-wave superconductor with $k_x\pm ik_y$ gap symmetry
can break up into domains of opposite chirality separated by a domain wall, in which the relative phase of the order parameter components rotates between $-\pi/2$ and $\pi/2$ (Ref. \onlinecite{chiral-DW}).

Previous studies of the phase solitons in multiband superconductors focused on finding the soliton shape and energy in the Ginzburg-Landau regime.\cite{Tanaka02,GV03,Vakar12,KOY11} In this paper, we investigate the effect of the phase solitons on the 
Bogoliubov quasiparticles. The presence of a nonuniform texture in the relative phase implies that the order parameter phases in individual bands also have kink-like inhomogeneities. We show that, in addition to the gapped quasiparticles in the bulk,
there are states in both bands which are localized near the soliton and have energies below the bulk gap edges. The origin of these states is similar to that of the fermion states bound to topological defects, which have appeared in many different contexts in 
high energy and condensed matter physics.\cite{bound-fermions} A different type of subgap states that can exist near the surface of a two-band superconductor of the $s_\pm$ symmetry, i.e. when the order parameters in the bands have opposite signs, was discussed
in Ref. \onlinecite{ABS-two-band}.

The paper is organized as follows: In Sec. \ref{sec: GL description}, we discuss the structure of the phase soliton in the Ginzburg-Landau regime and show how the kink in the relative phase is translated into kinks in the individual condensate phases, 
with non-universal phase winding numbers. In Sec. \ref{sec: QP spectrum}, we study the quasiparticle spectrum in the presence of a single planar soliton using semiclassical, or Andreev, equations and calculate the energy of the bound states. In Sec. \ref{sec: Energy}, 
the phase soliton energy is calculated using an exact representation of the functional determinant of the Andreev Hamiltonian. Throughout the paper we use the units in which $\hbar=k_B=1$.

\section{Ginzburg-Landau description of the phase soliton}
\label{sec: GL description}

We assume a clean superconductor with two isotropic bands, labeled by $a=1,2$, and isotropic $s$-wave singlet pairing, described by two order parameters $\eta_1(\br)$ and $\eta_2(\br)$, in the absence of a magnetic field. 
The difference between the free energies in the superconducting and normal states is given by ${\cal F}_s-{\cal F}_n=\int f_{GL}\,d^3\br$, where
\begin{equation}
\label{GL_gen}
  f_{GL}=\sum_{a}\left[\alpha_a|\eta_a|^2+\frac{\beta_a}{2}|\eta_a|^4+K_a|\bm{\nabla}\eta_a|^2\right]+\gamma(\eta_1^*\eta_2+\eta_2^*\eta_1).
\end{equation}
The intraband terms have the usual Ginzburg-Landau form, while the last term describes the interband ``Josephson coupling'', i.e. the Cooper pair tunneling between the bands. 

Using the amplitude-phase representation of the order parameter, $\eta_a(\br)=|\eta_a(\br)|e^{i\varphi_a(\br)}$, the free energy density can be written as
\begin{equation}
\label{GL-ap}
  f_{GL}=\sum_{a}\left[\alpha_a|\eta_a|^2+\frac{\beta_a}{2}|\eta_a|^4+K_a(\bm{\nabla}|\eta_a|)^2+K_a|\eta_a|^2(\bm{\nabla}\varphi_a)^2\right]+2\gamma|\eta_1||\eta_2|\cos(\varphi_1-\varphi_2).
\end{equation}
In a uniform state, the minimum energy corresponds to $|\eta_a|=\Delta_a$ and $\varphi_1-\varphi_2=\theta_0(\mathrm{mod}\,2\pi)$, where
\begin{equation}
\label{theta_0}
  \theta_0=0,\quad\mathrm{if\ }\gamma<0,\qquad\theta_0=\pi,\quad\mathrm{if\ }\gamma>0. 
\end{equation}
The first possibility (interband attraction) is realized in MgB$_2$, in which both gaps have the same phase,\cite{MgB2-gaps} while the second possibility (interband repulsion) is likely realized in the iron pnictides, in which, 
according to the most popular model, the gap function reverses its sign between different sheets of the Fermi surface, corresponding to the so-called $s_\pm$ pairing.\cite{s-pm}

It follows from Eq. (\ref{GL-ap}) that the supercurrent is a sum of independent contributions from individual bands: $\bm{j}=-(4e/c)\sum_aK_a|\eta_a|^2(\bm{\nabla}\varphi_a)$ ($e$ is the absolute value of electron charge). 
For a planar texture perpendicular to the $x$ axis, the current conservation implies that $\bm{j}=j\hat{\bm{x}}$, where $j$ is a constant. The value of the current is set by external sources and can be assumed to be zero.  
In order for the supercurrent contributions from bands 1 and 2 to cancel each other, the two order parameter phases must vary in a counterphase fashion, with $\nabla_x\varphi_2=-\rho(x)\nabla_x\varphi_1$, where 
$\rho=K_1|\eta_1|^2/K_2|\eta_2|^2$. This allows one to express the free energy (\ref{GL-ap}) in terms of the relative phase $\theta=\varphi_1-\varphi_2$: 
\begin{equation}
\label{f-GL-theta}
  f_{GL}=\sum_{a}\left[\alpha_a|\eta_a|^2+\frac{\beta_a}{2}|\eta_a|^4+K_a(\bm{\nabla}|\eta_a|)^2\right]+\frac{K_1K_2|\eta_1|^2|\eta_2|^2}{K_1|\eta_1|^2+K_2|\eta_2|^2}(\nabla_x\theta)^2+2\gamma|\eta_1||\eta_2|\cos\theta,
\end{equation}
Variational minimization of this expression yields a system of three coupled nonlinear differential equations for $|\eta_1(x)|$, $|\eta_2(x)|$, and $\theta(x)$,
with the asymptotics $|\eta_a(\pm\infty)|=\Delta_a$ and $\theta(\pm\infty)=\theta_0(\mathrm{mod}\,2\pi)$. 

In addition to the uniform solutions, the order parameter equations have various nonuniform ones, connecting different degenerate minima of $\cos\theta$. The simplest topologically nontrivial solutions are those with
$\theta(+\infty)-\theta(-\infty)=\pm 2\pi$, where the positive (negative) sign corresponds to a phase soliton (anti-soliton). 
The presence of a soliton texture in the relative phase implies that each of the two phases $\varphi_1$ and $\varphi_2$ is also spatially nonuniform and attains different values at $x=+\infty$ and $x=-\infty$. We define the phase winding parameter as
\begin{equation}
\label{chi_1}
  \chi\equiv\varphi_1(+\infty)-\varphi_1(-\infty)=\int_{-\infty}^{+\infty}dx\frac{\nabla_x\theta}{1+\rho(x)},
\end{equation}
then $\varphi_2(+\infty)-\varphi_2(-\infty)=\chi\mp 2\pi$ for the soliton (anti-soliton).

An explicit expression for the phase soliton can be obtained in the London approximation, when the order parameter amplitudes are constant everywhere, i.e. $|\eta_a(x)|=\Delta_a$ (Ref. \onlinecite{Tanaka02}). The minimization of Eq. (\ref{f-GL-theta}) 
then yields a static sine-Gordon equation for the relative phase, whose soliton solution has the form $\theta(x)=\theta_s(x)+(\pi-\theta_0)$, where $\theta_s(x)=2\arcsin[\tanh(x/\xi_s)]$ and
$$
  \xi_s=\sqrt{\frac{K_1K_2\Delta_1\Delta_2}{(K_1\Delta_1^2+K_2\Delta_2^2)|\gamma|}}
$$
has the meaning of the soliton width. The phase textures in the bands are given by the following expressions (up to a common phase rotation):
\begin{equation}
\label{phi-12-L}
  \varphi_1(x)=\frac{1}{1+\rho_0}\theta_s(x),\qquad \varphi_2(x)=-\frac{\rho_0}{1+\rho_0}\theta_s(x)-(\pi-\theta_0),
\end{equation}
where $\rho_0=K_1\Delta_1^2/K_2\Delta_2^2$. In the London approximation, the phase winding parameter, see Eq. (\ref{chi_1}), takes the form $\chi=2\pi/(1+\rho_0)$.

\section{Quasiparticle spectrum}
\label{sec: QP spectrum}

The qualitative features of the phase soliton discussed above are expected to survive beyond the Ginzburg-Landau regime. Namely, the phase soliton divides the superconductor into two domains, separated by a ``domain wall'', whose thickness is of the
order of $\xi_s$. 
The order parameter phase in each of the bands exhibits a kink-like variation, similar to the London-limit expressions, see Eq. (\ref{phi-12-L}), with $\varphi_a(+\infty)-\varphi_a(-\infty)=\chi_a$. For a single soliton, we have 
\begin{equation}
\label{chi-relation}
  \chi_1=\chi,\qquad \chi_2=\chi-2\pi
\end{equation}
where the phase winding parameter $\chi$ is a non-universal fraction of $2\pi$, determined by the microscopic details. 

Now we turn to the calculation of the quasiparticle spectrum in the presence of a single planar soliton. The bands are isotropic, with the dispersions $\xi_a(\bk)=(\bk^2-k_{F,a}^2)/2m_a$, characterized by the effective masses $m_a$ and 
the Fermi wave vectors $k_{F,a}$. Since the order parameters vary slowly on the atomic length scales, one can use the semiclassical, or Andreev, approximation.\cite{And64} An important point is that the slow perturbation due to the phase soliton 
cannot cause quasiparticle transitions between the bands, therefore one can solve the Andreev equations independently in each band. 

Quasiparticles propagating along the semiclassical trajectory directed along the unit vector $\hat{\bk}_F$ are described by the wave function $\psi$, which varies slowly compared to $k_F^{-1}$. The quasiparticle spectrum at given $\hat{\bk}_F$
is determined by the equation $\hat H\psi=E\psi$, where the Andreev Hamiltonian is given by
\begin{equation}
\label{And-eq}
	\hat H=\left(\begin{array}{cc}
		-iv_{F,x}\nabla_x & \eta(x)\\
		\eta^*(x) & iv_{F,x}\nabla_x
	\end{array}\right).
\end{equation}
Here $\bm{v}_F=\bk_F/m$ is the Fermi velocity and $\eta(x)=|\eta(x)|e^{i\varphi(x)}$. The gap magnitude approaches its bulk mean-field value far from the soliton: $|\eta(x)|\to\Delta_0$ at $|x|\gg\xi_s$ [the London approximation corresponds to 
$|\eta(x)|=\Delta_0$ everywhere]. While the band index has been temporarily dropped for brevity, we note that in the $a$th band, $\bm{v}_F\to\bm{v}_{F,a}=(k_{F,a}/m_a)\hat{\bk}_F$, $\Delta_0\to\Delta_a$, and $\varphi(x)\to\varphi_a(x)$. 

To make the eigenvalue problem for the Hamiltonian (\ref{And-eq}) well-defined, we put the system in a box of length $\ell$, such that $\ell\gg\xi_s$. When safe to do so, we will take the limit $\ell\to\infty$. 
For consistency with the phase winding of the order parameter, one should use twisted boundary conditions for the quasiparticle wave functions: $\psi(+\ell/2)=e^{i\chi\hat\sigma_3/2}\psi(-\ell/2)$. 

It is convenient to represent the phase soliton as a localized perturbation, which is achieved by applying a gauge transformation: $\psi=\hat U\tilde\psi$ and $\hat U^\dagger\hat HU=\hat{\tilde H}$, where 
\begin{equation}
\label{U-def}
  \hat U(x)=e^{i\varphi(x)\hat\sigma_3/2}.
\end{equation}
We can drop the tildas and write the transformed Hamiltonian in the form
\begin{equation}
\label{H-transformed}
  \hat H=\hat H_0+\delta\hat H,
\end{equation}
where 
\begin{equation}
\label{H_0}
  \hat H_0=-iv_{F,x}\hat\sigma_3\nabla_x+\Delta_0\hat\sigma_1
\end{equation}
describes the Bogoliubov quasiparticles in the uniform superconducting state, while
$$
  \delta\hat H=\frac{1}{2}v_{F,x}\varphi'(x)\hat\sigma_0+\left[|\eta(x)|-\Delta_0\right]\hat\sigma_1
$$
represents a perturbation which is nonzero only near the soliton, i.e. at $|x|\lesssim\xi_s$. The gauge-transformed eigenfunctions satisfy the periodic boundary conditions: 
\begin{equation}
\label{periodic-bc}
  \psi\left(+\frac{\ell}{2}\right)=\psi\left(-\frac{\ell}{2}\right).
\end{equation}
Note that there is a one-to-one correspondence between the spectra of $\hat H$ and $\hat H_0$: the eigenvalues of the operator $\hat H_s=\hat H_0+\lambda\delta\hat H$ evolve smoothly between those of $\hat H_0$ and $\hat H$ as 
the parameter $\lambda$ varies between $0$ and $1$.

At given $\hat{\bk}_F$, the spectrum of the Andreev Hamiltonian consists of scattering states with the energies $|E|\geq\Delta_0$ and bound states with $|E|<\Delta_0$. 
Let us start with the former. Far from the soliton, the Hamiltonian is equal to $\hat H_0$ and the scattering eigenstates are the superpositions of plane waves:
\begin{equation}
\label{psi-scatter}
  \psi(x)\bigr|_{x\to\pm\infty}=C^{\pm}_R\left(\begin{array}{c}
                           w_R \\ 1 
                           \end{array}\right)e^{iqx}+
	  C^{\pm}_L\left(\begin{array}{c}
                           w_L \\ 1 
                           \end{array}\right)e^{-iqx},
\end{equation}
where $q=\sqrt{E^2-\Delta_0^2}/|v_{F,x}|>0$, $w_{R(L)}=\Delta_0/(E\mp v_{F,x}q)$, and the subscripts $R,L$ refer to the direction of propagation of the corresponding waves. 
The coefficients in these asymptotics are not independent: it is convenient to introduce a $2\times 2$ scattering matrix, or the $S$-matrix, which expresses the amplitudes of the outgoing waves in terms of the amplitudes of the incoming waves:
\begin{equation}
\label{S-matrix}
  \left(\begin{array}{c} C^{+}_R \smallskip \\ C^{-}_L \end{array}\right)=\hat S\left(\begin{array}{c} C^{-}_R \smallskip \\ C^{+}_L \end{array}\right).
\end{equation}
The elements of the $S$-matrix depend on the energy and are determined by the details of the order parameter at $|x|\lesssim\xi_s$.
Note that the $S$-matrix defined by Eq. (\ref{S-matrix}) is not unitary, in general, since we did not bother to normalize the scattering states. Still, one can show that the $S$-matrix satisfies a certain constraint, which follows
from a ``conservation law'' for the Andreev equations. It is straightforward to check that $\nabla_x(\psi^\dagger\hat\sigma_3\psi)=0$ for the eigenfunctions of Eq. (\ref{And-eq}), therefore,
$\psi^\dagger(x)\hat\sigma_3\psi(x)=\mathrm{const}$. Substituting here the asymptotical expressions (\ref{psi-scatter}) and using the definition (\ref{S-matrix}), we obtain that the $S$-matrix must satisfy $\hat S^\dagger\hat\mu\hat S=\hat\mu$, 
where $\hat\mu=\mathrm{diag}(w_R^2-1,1-w_L^2)$. In particular, $|\det\hat S|=1$. 

One can also introduce the $\tau$-matrix, which relates the scattering wave amplitudes at $x\to+\infty$ to those at $x\to-\infty$:
\begin{equation}
\label{tau-def}
  \left(\begin{array}{c} C^{+}_R \smallskip \\ C^{+}_L \end{array}\right)=\hat\tau\left(\begin{array}{c} C^{-}_R \smallskip \\ C^{-}_L \end{array}\right).
\end{equation}
Comparing Eqs. (\ref{tau-def}) and (\ref{S-matrix}), we find that the $\tau$-matrix can be expressed in terms of the $S$-matrix: 
\begin{equation}
\label{tau-S}
  \hat\tau=\frac{1}{S_{22}}\left(\begin{array}{cc}
                   \det\hat S & S_{12} \smallskip \\
		   -S_{21} & 1 
                   \end{array}\right).
\end{equation}
In the absence of the phase soliton, there is no scattering and $\hat S=\hat\tau=\hat\sigma_0$.

\subsection{Subgap bound states}
\label{sec: subgap BS}

The $S$-matrix (or the $\tau$-matrix) can also be used to obtain the bound states, which correspond to the poles at $|E|<\Delta_0$ on the real axis in the complex energy plane. The function $q(z)=\sqrt{z^2-\Delta_0^2}/|v_{F,x}|$, 
where $z$ is the complex energy, has two branch points at $z=\pm\Delta_0$. The appropriate branch of $q(z)$ is fixed by the condition that, as implied by Eq. (\ref{psi-scatter}), $q$ is a positive real number when $z$ is outside the gap on the real axis,
i.e. when $z=E$ with $|E|>\Delta_0$. 
One can select the branch cuts to run parallel to the imaginary axis, from $\pm\Delta_0$ to $\pm\Delta_0\mp i\infty$. Then, at $|E|<\Delta_0$ we have $q=i\Omega/|v_{F,x}|$, where $\Omega=\sqrt{\Delta_0^2-E^2}$.

The $S$-matrix can be calculated analytically in a simple model, in which the soliton width is sent to zero, so that 
\begin{equation}
\label{sharp-soliton-def}
  |\eta(x)|=\Delta_0,\quad \varphi(x<0)=0,\quad \varphi(x>0)=\chi,
\end{equation}
where $\chi$ is the phase winding parameter. The gauge transformation operator $\hat U(x)$, see 
Eq. (\ref{U-def}), is discontinuous at $x=0$, which implies the following matching condition for the gauge-transformed wave function: $\psi(+0)=e^{-i\chi\hat\sigma_3/2}\psi(-0)$. After a straightforward calculation, we obtain:
\begin{equation}
\label{S-sharp}
  \hat S=\left(\cos\frac{\chi}{2}+i\frac{E}{v_{F,x}q}\sin\frac{\chi}{2}\right)^{-1}\left(\begin{array}{cc}
              1 & i\left(1-\dfrac{E}{v_{F,x}q}\right)\sin\dfrac{\chi}{2} \\ 
              -i\left(1+\dfrac{E}{v_{F,x}q}\right)\sin\dfrac{\chi}{2} & 1
              \end{array}\right).
\end{equation}
The characteristic equation for the bound states at $|E|<\Delta_0$ has the form 
\begin{equation}
\label{char-eq-sharp}
  \cos\frac{\chi}{2}+\sgn(v_{F,x})\frac{E}{\Omega}\sin\frac{\chi}{2}=0.
\end{equation}
Introducing $\tilde E=E\sgn(v_{F,x})$, one can write $\tilde E=\Delta_0\cos\Theta$ and $\Omega=\Delta_0\sin\Theta$. 
Since, according to Eq. (\ref{char-eq-sharp}), $\tan\Theta=-\tan(\chi/2)$, we have $\Theta=-\chi/2+\pi n$ ($n$ is an integer) and, therefore, $\tilde E=\Delta_0(-1)^n\cos(\chi/2)$. The parity of $n$ can be found from the condition $\Omega\geq 0$, which yields 
$(-1)^n=-\sgn\left[\sin(\chi/2)\right]$. Collecting everything together, we obtain:
\begin{equation}
  E=-\Delta_0\sgn\left(v_{F,x}\sin\frac{\chi}{2}\right)\cos\frac{\chi}{2},
\end{equation}
i.e. there is a single bound state with the energy inside the bulk gap. In the absence of the soliton, i.e. at $\chi=0$, we have $|E|=\Delta_0$, i.e. the bound state merges into the continuum of the bulk states. 
Note that the ``sharp'' phase soliton is formally similar to a Josephson junction between two $s$-wave superconductors, with the phase difference equal to $\chi$. The bound state energy for such a junction was calculated in Ref. \onlinecite{s-wave-subgap}. 

Restoring the band indices and using the phase windings from Eq. (\ref{chi-relation}), we finally obtain that there is one subgap bound state for each direction of semiclassical propagation $\hat{\bk}_F$ in each of the bands, with the energy given by
\begin{equation}
\label{E-bound}
  E_a=-\Delta_a\sgn(v_{F,a,x})\cos\frac{\chi}{2}.
\end{equation}
We see that the bound state energy is a non-universal fraction of the bulk gap. It is only in the exceptional case when the microscopic parameters are fine tuned to yield $\chi=\pi$, that the subgap states are located exactly at zero energy.  

Expression (\ref{E-bound}) has the property $E_a(-\hat{\bk}_F)=-E_a(\hat{\bk}_F)$, which is a consequence of the ``electron-hole'' symmetry of the Andreev spectrum: for any $\hat{\bk}_F$, if $\psi_{\hat{\bk}_F}$ is an eigenfunction of
the Andreev Hamiltonian $\hat H_{\hat{\bk}_F}$ corresponding to the eigenvalue $E$, then $i\hat\sigma_2\psi^*_{\hat{\bk}_F}$ is an eigenfunction of the Andreev Hamiltonian $\hat H_{-\hat{\bk}_F}$ corresponding to the eigenvalue $-E$.
After angular averaging over the Fermi surface, the bound states will manifest themselves as four $\delta$-function peaks in the quasiparticle density of states, located symmetrically at $E=\pm\Delta_{1,2}\cos(\chi/2)$.

\section{Energy of the phase soliton}
\label{sec: Energy}

Since the quasiparticles bound to the phase soliton have lower energies than in the bulk, it is natural to ask whether the spontaneous formation of solitons, accompanied by ``self-trapping'' of quasiparticles, could be possible. The general expression 
for the energy of a nonuniform state in a two-band superconductor is derived in the Appendix. For a planar order parameter texture with $\eta_a(x)=|\eta_a(x)|e^{i\varphi_a(x)}$, in particular, for the phase soliton, the free energy difference per unit area 
between the states with and without the soliton has the form $F_s=\delta{\cal F}/A_\perp$, where $\delta{\cal F}$ is given by Eq. (\ref{delta-F-gen}) and $A_\perp$ is the area of the system in the directions perpendicular to $x$.
We have $F_s=F_1+F_2$, where 
\begin{equation}
\label{F_1-def}
  F_1=-T\sum_n\sum_a\int\frac{d^2\bm{k}_\perp}{(2\pi)^2}\sum_\mu\ln\frac{i\omega_n-E_{a,\bm{k}_\perp,\mu}}{i\omega_n-E^{(0)}_{a,\bm{k}_\perp,\mu}}
\end{equation}
and
\begin{equation}
\label{F_2-def}
  F_2=\int dx\sum_{ab}(\hat V^{-1})_{ab}\left(\eta_a^*\eta_b-\eta^*_{a,0}\eta_{b,0}\right).
\end{equation}
In $F_1$, we used the following notations: $\omega_n=(2n+1)\pi T$ is the fermionic Matsubara frequency, $\bm{k}_\perp=(k_y,k_z)$ is the wave vector parallel to the soliton, and $\mu$ labels the eigenstates of the reduced one-dimensional 
BdG Hamiltonian in the $a$th band, at given $\bk_\perp$:
\begin{equation}
\label{H-BdG-perp}
  \hat H^{BdG}_{a,\bm{k}_\perp}=\left(\begin{array}{cc}
                                \dfrac{\hat k_x^2-k_{0,a}^2}{2m_a} & \eta_a(x) \\
				\eta_a^*(x) & -\dfrac{\hat k_x^2-k_{0,a}^2}{2m_a}
                                \end{array}\right),
\end{equation}
where $k_{0,a}=\sqrt{k_{F,a}^2-k_\perp^2}$. In $F_2$, $V_{ab}$ are the coupling constants of the intraband and interband pairing, see Appendix for the explanation.

Since $F_1$ is a sum of the independent contributions from the two bands, we can drop the band index temporarily. 
As in Sec. \ref{sec: QP spectrum}, one can use the Andreev approximation to find the spectrum of the Hamiltonian (\ref{H-BdG-perp}), because the order parameter varies slowly on the scale of the inverse Fermi wave vector.
We seek the eigenfunctions of Eq. (\ref{H-BdG-perp}) in the form $\Psi(x)=e^{ik_xx}\psi(x)$, where $k_x=\pm k_0$. The direction of semiclassical propagation of quasiparticles is defined by the wave vector 
$\bk_F\equiv (\bm{k}_\perp,k_x)=k_F\hat{\bk}_F$. The slowly-varying function $\psi(x)$ is found by solving the eigenvalue equation $\hat H\psi=E\psi$, where $\hat H$ is the Andreev Hamiltonian, see Eq. (\ref{And-eq}). 
The sum over the BdG spectrum in Eq. (\ref{F_1-def}) can be expressed in the semiclassical approximation in terms of a Fermi-surface angular average of a sum over the Andreev spectrum:
$$
  \int\frac{d^2\bm{k}_\perp}{(2\pi)^2}\sum_\mu(...)=2\pi N_F\int\frac{d\hat{\bk}_F}{4\pi}|v_{F,x}|\sum_i(...),
$$
where $N_F=mk_F/2\pi^2$ is the Fermi-level density of states and $i$ labels the eigenstates of the Andreev Hamiltonian at given $\hat{\bk}_F$.
 
Removing the order parameter phase by the gauge transformation (\ref{U-def}) and restoring the band indices, we finally arrive at the following result:
\begin{equation}
\label{F_1-final-Andreev}
  F_1=-2\pi T\sum_n\sum_a N_{F,a}\int\frac{d\hat{\bk}_F}{4\pi}|v_{F,a,x}|\,\ln D_{a,\hat{\bk}_F}(i\omega_n),
\end{equation}
where
\begin{equation}
\label{Andreev-det}
  D_{a,\hat{\bk}_F}(z)=\prod_i\frac{z-E_i(a,\hat{\bk}_F)}{z-E^{(0)}_i(a,\hat{\bk}_F)}=\frac{\det[z-\hat H(a,\hat{\bk}_F)]}{\det[z-\hat H_0(a,\hat{\bk}_F)]}
\end{equation}
is the ratio of the functional determinants of the Andreev Hamiltonians in the nonuniform and uniform states, see Eqs. (\ref{H-transformed}) and (\ref{H_0}), for a given direction of the semiclassical propagation on the Fermi surface in the $a$th band.

\subsection{Calculation of the functional determinant}
\label{sec: calculation of D}

There exists a very efficient way of calculating the expression (\ref{Andreev-det}), which is based on a relation between the functional determinant and the transfer matrix for the Andreev Hamiltonian (see Ref. \onlinecite{Wax94}, 
where a closely related Dirac Hamiltonian was investigated). Let us again drop the band and direction indices, $a$ and $\hat{\bk}_F$. 
The transfer matrix is defined as a $2\times 2$ matrix satisfying the equation $(z-\hat H)\hat M(x;z)=0$, where $\hat H$ is given by Eq. (\ref{H-transformed}), with the initial condition $\hat M(-\ell/2;z)=\hat\sigma_0$.
Since the eigenfunctions of $\hat H$ can be written as $\psi(x)=\hat M(x;z)\psi(-\ell/2)$, the transfer matrix has the meaning of the evolution operator of the wave functions along the $x$-axis.

From the periodic boundary condition (\ref{periodic-bc}) we obtain the characteristic equation for the eigenvalues: $\det[\hat\sigma_0-\hat m(z)]=0$, where we introduced a shorthand notation, $\hat m(z)\equiv\hat M(\ell/2;z)$,
for the transfer matrix from one end of the system to the other. The ratio of the functional determinants, Eq. (\ref{Andreev-det}), can then be represented in the following form:
\begin{equation}
\label{D-gen-m}
  D(z)=\frac{\det[\hat\sigma_0-\hat m(z)]}{\det[\hat\sigma_0-\hat m_0(z)]},
\end{equation}
where $\hat m_0$ is the transfer matrix from $-\ell/2$ to $\ell/2$ for $\hat H_0$. That the two sides of Eq. (\ref{D-gen-m}) have to be the same immediately follows from the fact that they both are meromorphic functions in the complex energy plane, 
having the same poles and zeros and also the same asymptotics at $|z|\gg\Delta_0$. If this argument is not convincing, a more elaborate proof can be found in Ref. \onlinecite{Wax94}. 
An expression like Eq. (\ref{D-gen-m}) represents a significant step forward compared to the definition (\ref{Andreev-det}), because it reduces the calculation of the infinitely-dimensional functional determinant to solving an initial value problem
for a $2\times 2$ transfer matrix. Expressions of this sort are sometimes called the Gelfand-Yaglom formulas, see Ref. \onlinecite{GY60} and also Ref. \onlinecite{Dunne08} for a review. 

Further simplification is possible in the thermodynamic limit, $\ell\to\infty$, where one can
represent Eq. (\ref{D-gen-m}) in terms of the scattering matrix. To obtain this representation, we note that, according to the definition of the transfer matrix, 
\begin{equation}
\label{psi-pm}
  \psi\left(+\frac{\ell}{2}\right)=\hat m\psi\left(-\frac{\ell}{2}\right).
\end{equation}
On the other hand, using Eq. (\ref{psi-scatter}), the wave functions far from the phase soliton can be expressed in terms of the scattering wave amplitudes as follows:
\begin{equation}
\label{W-pm-def}
  \psi\left(\pm\frac{\ell}{2}\right)=\hat W_\pm\left(\begin{array}{c}
                                   C^{\pm}_R\smallskip \\ C^{\pm}_L
                                   \end{array}\right),
\end{equation}
where
$$
  \hat W_\pm=\left(\begin{array}{cc}
                   w_Re^{\pm iq\ell/2} & w_Le^{\mp iq\ell/2} \\
		   e^{\pm iq\ell/2} & e^{\mp iq\ell/2} 
                   \end{array}\right).
$$
It follows from Eqs. (\ref{tau-def}), (\ref{psi-pm}), and (\ref{W-pm-def}) that $\hat m=\hat W_+\hat\tau\hat W^{-1}_-$ and
\begin{equation}
\label{D-gen-tau}
  D(z)=\frac{\det(\hat\sigma_0-\hat W_+\hat\tau\hat W^{-1}_-)}{\det(\hat\sigma_0-\hat W_+\hat W^{-1}_-)}.
\end{equation}
Here we used the fact that $\hat\tau=\hat\sigma_0$ for $\hat H_0$. 

According to Eq. (\ref{F_1-final-Andreev}), the free energy of the phase soliton is expressed in terms of the Andreev functional determinant on the imaginary energy axis. 
At $z=i\omega_n$, we have $q=i\kappa$, where $\kappa=\sqrt{\omega_n^2+\Delta_0^2}/|v_{F,x}|$. Calculating the $2\times 2$ determinants on the right-hand side of Eq. (\ref{D-gen-tau}) and keeping only the leading, exponentially divergent at 
$\ell\to\infty$, terms, we obtain: $\det(\hat\sigma_0-\hat W_+\hat\tau\hat W^{-1}_-)=-e^{\kappa\ell}\tau_{22}$ and $\det(\hat\sigma_0-\hat W_+\hat W^{-1}_-)=-e^{\kappa\ell}$. Therefore, 
\begin{equation}
\label{D-Matsubara}
  D(i\omega_n)\bigr|_{\ell\to\infty}=\tau_{22}(i\omega_n)=\frac{1}{S_{22}(i\omega_n)},
\end{equation}
where we used the relation (\ref{tau-S}) between the $\tau$- and $S$-matrices.

Returning to Eq. (\ref{F_1-final-Andreev}), we finally obtain:
\begin{equation}
\label{F_1-final-S}
  F_1=2\pi T\sum_n\sum_a N_{F,a}\int\frac{d\hat{\bk}_F}{4\pi}|v_{F,a,x}|\,\ln S_{22}(i\omega_n;a,\hat{\bk}_F).
\end{equation}
Thus, the problem of evaluating the free energy of a nonuniform order parameter texture has been reduced to the calculation of the semiclassical scattering matrix of the Bogoliubov quasiparticles, analytically continued to complex energies.

\subsection{Sharp soliton}
\label{sec: Sharp soliton energy}

The scattering matrix can be calculated explicitly only in some simple cases. For instance, for the sharp phase soliton defined in Sec. \ref{sec: subgap BS}, it is given by Eq. (\ref{S-sharp}) and we have
$$
  T\sum_n\ln S_{22}(i\omega_n;a,\hat{\bk}_F)=-T\sum_{n\geq 0}\ln\left(1-\frac{\Delta_a^2}{\omega_n^2+\Delta_a^2}\sin^2\frac{\chi}{2}\right).
$$
According to Eq. (\ref{F_2-def}), for the sharp soliton $F_2$ vanishes and we obtain the following exact expression for the energy, which is valid at all temperatures:
\begin{equation}
\label{F sharp soliton}
  F_s=-\pi\sum_aN_{F,a}v_{F,a}\;T\sum_{n\geq 0}\ln\left(1-\frac{\Delta_a^2}{\omega_n^2+\Delta_a^2}\sin^2\frac{\chi}{2}\right).
\end{equation}
At $T=0$, the Matsubara sum here becomes an integral and can be calculated in a closed form:
$$
  \int_0^\infty\frac{d\omega}{2\pi}\ln\left(1-\frac{\Delta_a^2}{\omega^2+\Delta_a^2}\sin^2\frac{\chi}{2}\right)=-\frac{\Delta_a}{2}\left(1-\left|\cos\frac{\chi}{2}\right|\right).
$$
Therefore,
\begin{equation}
\label{F sharp soliton zero t}
  F_s(T=0)=\frac{\pi}{2}\left(1-\left|\cos\frac{\chi}{2}\right|\right)\sum_aN_{F,a}v_{F,a}\Delta_a.
\end{equation}

Expressions (\ref{F sharp soliton}) and (\ref{F sharp soliton zero t}) show that the soliton energy is positive, vanishing only in the absence of the phase winding, i.e. at $\chi=0$. 
Thus we come to the conclusion that the spontaneous formation of the phase solitons is energetically unfavorable. Note though that a definitive answer would require a self-consistent solution of the gap equations. 
The feedback effect of the subgap states on the order parameter profile might be strong enough to cause
self-trapping of Bogoliubov quasiparticles, similar to that discussed in Ref. \onlinecite{KY02}, see also Ref. \onlinecite{self-trapped}. Investigation of this possibility is beyond the scope of the present work.

\section{Conclusions}
\label{sec: Conclusion}

We studied the Bogoliubov quasiparticle spectrum in a two-band superconductor, in the presence of a soliton-like topological defect in the relative phase $\varphi_1-\varphi_2$. 
While the relative phase winding across the soliton is given by $2\pi$, the phase windings in individual bands are non-universal fractions of $2\pi$: $\varphi_1(+\infty)-\varphi_1(-\infty)=\chi$ and $\varphi_2(+\infty)-\varphi_2(-\infty)=\chi-2\pi$, 
where the parameter $\chi$ depends on the microscopic details. We found that there are quasiparticle bound states localized near the soliton, whose 
energies are non-universal fractions of the bulk gaps. 

The bound states will lead to sharp peaks in the quasiparticle density of states at $E=\pm\Delta_{1,2}\cos(\chi/2)$, which can be 
observed in tunneling experiments. The tunneling probe will have to be located sufficiently close to the phase soliton to be able to detect the contribution from the localized states. This can be done, e.g. in the experimental setup proposed in 
Ref. \onlinecite{Vakar12}, in which the soliton is ``pinned'' to the spatial variation of the interband Josephson coupling, controlled by the proximity effect with another superconductor.
We note that the peaks in the density of states are expected to acquire a finite width when impurity scattering or intraband gap anisotropy are taken into account. 

We also derived a general expression for the phase soliton energy, relating it to the scattering matrix of the Bogoliubov quasiparticles. As a simple application, we exactly calculated the energy in the limit of zero soliton width.

\acknowledgments

This work was supported by a Discovery Grant from the Natural Sciences and Engineering Research Council (NSERC) of Canada.

\appendix

\section{Free energy of a nonuniform two-band superconductor}
\label{app: Microscopics}

In this Appendix, we present a microscopic derivation of the free energy of a clean two-band superconductor in a nonuniform state, at arbitrary temperature, using the effective action formalism. We start with a two-band generalization of the BCS Hamiltonian:
\begin{equation}
\label{H-two-band}
  \hat{\cal H}=\sum_{\bk,a}\xi_a(\bk)\hat c^\dagger_{\bk,a,\alpha}\hat c_{\bk,a,\alpha}-\frac{1}{\cal V}\sum_{\bk\bk'\bq,ab}V_{ab}\hat c^\dagger_{\bk+\bq,a,\uparrow}\hat c^\dagger_{-\bk,a,\downarrow}
    \hat c_{-\bk',b,\downarrow}\hat c_{\bk'+\bq,b,\uparrow},
\end{equation}
where $a=1,2$ is the band index, $\alpha=\uparrow,\downarrow$ is the spin projection (the spin indices that appear twice are summed over), and ${\cal V}$ is the system volume. 
The second term in the Hamiltonian describes singlet $s$-wave pairing interactions in the Cooper channel: the intraband pairing, characterized by the coupling constants $V_{11}$ and $V_{22}$, and the interband ``tunneling'' of the pairs, 
described by $V_{12}$ and $V_{21}$. The Hermiticity and time-reversal invariance of the Hamiltonian dictate that the constants $V_{ab}$ form a real symmetric matrix $\hat V$. 

Our derivation is a straightforward generalization of the standard textbook procedure in the single-band case, see, e.g., Ref. \onlinecite{Pop91}. 
The partition function for the Hamiltonian (\ref{H-two-band}) can be represented as a functional integral over the Grassmann fields $c_{\bk,a,\alpha}(\tau)$ and $\bar c_{\bk,a,\alpha}(\tau)$:
$Z=\int{\cal D}c{\cal D}\bar c\;e^{-S[\bar c,c]}$, where the action is given by 
$$
  S=\int_0^\beta d\tau\sum_{\bk,a}\bar c_{\bk,a,\alpha}\frac{\partial}{\partial\tau} c_{\bk,a,\alpha}+\int_0^\beta d\tau\,{\cal H}[\bar c,c],
$$
with $\beta=1/T$. The interaction term in the action can be written as
$$
    S_{int}=-\int_0^\beta d\tau\;{\cal V}\sum_{\bq,ab}V_{ab}\bar B_a(\bq,\tau)B_b(\bq,\tau),
$$
where
$$
    \bar B_a(\bq,\tau)=\frac{1}{\cal V}\sum_{\bk}\bar c_{\bk+\bq,a,\uparrow}(\tau)\bar c_{-\bk,a,\downarrow}(\tau),\quad B_a(\bq,\tau)=\frac{1}{\cal V}\sum_{\bk}c_{-\bk,a,\downarrow}(\tau)c_{\bk+\bq,a,\uparrow}(\tau).
$$
One can decouple the interaction by means of the Hubbard-Stratonovich transformation, introducing two complex conjugated bosonic fields $\eta_{1,2}(\bq,\tau)$:
$$
    e^{-S_{int}}\to\int{\cal D}\eta^*{\cal D}\eta\;\exp\left\{-\int_0^\beta d\tau\,\frac{1}{\cal V}\sum_{\bq,ab}(\hat V^{-1})_{ab}\eta_a^*\eta_b
    -\int_0^\beta d\tau\sum_{\bq,a}(\eta_a^*B_a+\bar B_a\eta_a)\right\}.
$$
The field $\eta_a$ has the meaning of the fluctuating order parameter in the $a$th band. 

One can now calculate the Gaussian integral over the fermionic fields, to obtain $Z=\int{\cal D}\eta^*{\cal D}\eta\;e^{-S_{eff}[\eta^*,\eta]}$, where
\begin{equation}
\label{S-eff}
  S_{eff}=-\sum_a\Tr\ln\hat{G}_a^{-1}+\int_0^\beta d\tau\int d^3\br\sum_{ab}(\hat V^{-1})_{ab}\eta_a^*\eta_b
\end{equation}
is the effective bosonic action and
$$
  \hat{G}_a^{-1}=\left(\begin{array}{cc}
                                -\partial_\tau-\xi_a(\hat{\bk}) & -\eta_a(\br,\tau) \\
				-\eta^*_a(\br,\tau) & -\partial_\tau+\xi_a(\hat{\bk})
                                \end{array}\right)
$$
is the inverse Green's operator in the $a$th band, with $\hat{\bk}=-i\bm{\nabla}$. The trace in the first term should be understood as an operator trace in $(\br\tau)$-space and a $2\times 2$ matrix trace with respect to the electron-hole (Nambu) 
indices in the $a$th band.
Near the critical temperature, the order parameter components are small and the first term in Eq. (\ref{S-eff}) can be expanded in powers of $\eta_a$. In this way one would arrive at the Ginzburg-Landau functional for the two-band superconductor, 
which has been derived by different means in Ref. \onlinecite{two-band-GL}. 

We do not restrict ourselves to the Ginzburg-Landau regime and calculate the free energy in the mean-field approximation at arbitrary temperature.
The mean-field solution for the order parameter corresponds to a static saddle point of the effective action and satisfies the equations $\delta S/\delta\eta_{1,2}^*=0$. From Eq. (\ref{S-eff}) we obtain two coupled self-consistency equations:
\begin{equation}
\label{gap-eqs}
  \eta_a(\br)=-T\sum_n\sum_{b}V_{ab}G_{b,12}(\br,\br;\omega_n).
\end{equation}
Here $\omega_n=(2n+1)\pi T$ is the fermionic Matsubara frequency and $G_{a,12}$ is the anomalous (Gor'kov) component of the matrix Green's function $\hat{G}_a$, which satisfies the equation 
$(i\omega_n-\hat H^{BdG}_a)\hat{G}_a(\br,\br';\omega_n)=\hat\sigma_0\delta(\br-\br')$, where
\begin{equation}
\label{H-BdG}
  \hat H^{BdG}_a=\left(\begin{array}{cc}
                                \xi_a(\hat{\bk}) & \eta_a(\br) \\
				\eta^*_a(\br) & -\xi_a(\hat{\bk})
                                \end{array}\right)
\end{equation}
is the Bogoliubov-de Gennes (BdG) Hamiltonian.

The gap equations can be represented in terms of the eigenstates and eigenvalues of the BdG Hamiltonian, which are found from
$\hat H^{BdG}_a\Psi_{a,p}(\br)=E_{a,p}\Psi_{a,p}(\br)$. The eigenstates are two-component Nambu spinors, $\Psi=(u,v)^T$, labeled in the $a$th band by quantum numbers $p$. 
Assuming that the eigenstates form a complete and orthonormal set, we obtain for the matrix Green's function:
$$
  \hat{G}_a(\br,\br';\omega_n)=\sum_p\frac{\Psi_{a,p}(\br)\Psi^\dagger_{a,p}(\br')}{i\omega_n-E_{a,p}}.
$$
Inserting this into Eq. (\ref{gap-eqs}) and using the ``electron-hole'' symmetry of the BdG spectrum (if $\Psi$ corresponds to the energy $E$, then $i\hat\sigma_2\Psi^*$ corresponds to the energy $-E$), we arrive at the final form
of the gap equations:
\begin{equation}
\label{gap-eqs-final}
  \eta_a(\br)=\sum_bV_{ab}\sum_p{}'u_{b,p}(\br)v^*_{b,p}(\br)[1-2f(E_{b,p})],
\end{equation}
where $f(E)=1/(e^{\beta E}+1)$ is the Fermi function. The prime means that the summation is performed only over the upper half of the BdG spectrum, i.e. over the eigenstates with $E_{b,p}\geq 0$.
Eq. (\ref{gap-eqs-final}) can be used to obtain both the critical temperature and the temperature dependence of the gaps, see Refs. \onlinecite{two-band-BCS} and \onlinecite{two-band-GL}.
In addition to the spatially uniform solution, given by $\eta_{1,0}=\Delta_1e^{i\theta_0}$, $\eta_{2,0}=\Delta_2$, where $\theta_0=0$ for interband attraction ($V_{12}>0$) and $\theta_0=\pi$ for interband repulsion ($V_{12}<0$),
the gap equations also have various nonuniform solutions, in particular, the one corresponding to the phase soliton. 
 
The effective action (\ref{S-eff}) for any mean-field configuration of the order parameter has the form $S_{eff}=\beta{\cal E}$, where
$$
  {\cal E}=-T\sum_n\sum_{a,p}\ln(i\omega_n-E_{a,p})+\int d^3\br\sum_{ab}(\hat V^{-1})_{ab}\eta_a^*(\br)\eta_b(\br).
$$
The mean-field free energy is given by ${\cal F}=-T\ln Z=\mathrm{const}+{\cal E}$. To remove the undetermined constant, we calculate the free energy difference between a given nonuniform superconducting state and some reference state. For our
purposes, it is natural to choose the latter to be a uniform superconducting state with the order parameters equal to $\eta_{a,0}$, and we finally obtain:
\begin{equation}
\label{delta-F-gen}
    \delta{\cal F}\equiv{\cal F}[\eta]-{\cal F}[\eta_0]=-T\sum_n\sum_{a,p}\ln\frac{i\omega_n-E_{a,p}}{i\omega_n-E^{(0)}_{a,p}}+\int d^3\br\sum_{ab}(\hat V^{-1})_{ab}\left(\eta_a^*\eta_b-\eta^*_{a,0}\eta_{b,0}\right).
\end{equation}
Here $E^{(0)}_{a,p}$ are the eigenvalues of the BdG Hamiltonian (\ref{H-BdG}) in the uniform state.

\end{document}